\documentclass[12pt]{article}
\usepackage{scicite}
\usepackage{graphicx}
\usepackage{graphics}
\usepackage{amsfonts}
\usepackage{amsmath}
\usepackage{epsfig}
\usepackage{times}

\newenvironment{sciabstract}{%
\begin{quote} \bf}
{\end{quote}}

\newcounter{lastnote}

\title{\Large Comment on "Pairing and Phase Separation in a Polarized Fermi Gas" by G.~B. Partridge, W.~Li, R.~I. Kamar, Y.~Liao, R.~G. Hulet, {\it Science }{\bf 311}, 503 (2006)}

\author{Martin W. Zwierlein and Wolfgang Ketterle\\
\\
\normalsize{Department of Physics, MIT-Harvard Center for Ultracold Atoms,}\\
\normalsize{and Research Laboratory of Electronics, MIT, Cambridge, MA 02139}\\
\\
\normalsize{submitted: March 20, 2006;  revised: July 25, 2006}
}
\date{}
\begin{document}
\newcommand{\vect}[1]{\mathbf #1}

\maketitle

\begin{sciabstract}
We argue that it is not possible to infer from the results of
Partridge et al. which of their data was taken in the superfluid
or normal regime, and which of their clouds are phase-separated
and which are not.  Some of the conclusions in this paper are
inconsistent with recent experiments.
\end{sciabstract}

The authors of~\cite{part06phase} report the observation of a
quantum phase transition in a strongly interacting Fermi gas with
imbalanced spin populations. Below a critical population
imbalance, the paper claims evidence for a homogeneous superfluid
state with unequal densities.  Above the critical population, it
reports the observation of a core of superfluid pairs surrounded
by a shell of excess spin-up atoms.  In this comment we argue that
(a) the data presented in ~\cite{part06phase} do not allow to
unambiguously distinguish phase-separated from non phase-separated
clouds, and superfluid from normal clouds, (b) the paper does not
provide evidence for a quantum phase transition and (c) that the
claim of a quantum phase transition at small polarization and of
phase-separation and superfluidity at large polarization are
inconsistent with more detailed recent experiments.

It appears that the analysis in~\cite{part06phase} was carried out
without awareness that after double integration, a shell structure
does not lead to a density profile with two peaks at the edges,
but to a flat top-hat distribution since the density profile of a
thin hollow sphere after two integrations is flat. This was
already emphasized by~\cite{desi06dens}, who showed that any spin
distribution confined in a harmonic trap (with
$n_\uparrow(\vect{r}) > n_\downarrow(\vect{r})$ and assuming local
density approximation, LDA) results in flat~\cite{haqu06dens} or
monotonically decreasing difference profiles. The single peak
structures found by the authors of~\cite{part06phase} at small
population imbalances are thus the expected outcome for any state
of the gas, including phase-separated states.

Therefore, a question raised by the experiment
of~\cite{part06phase} is the origin of the double peak structures.
The double peaks require a breakdown of either the LDA (sometimes
called finite size effects) or the harmonic approximation.  Since
the LDA was generally expected to be valid for Fermi energies much
greater than the energy level spacing of the trap, we argued in
the first version of this comment that anharmonicities were a
plausible explanation~\cite{zwie06_comment}, given that
anharmonicities were not controlled or specified in the original
paper.  However, in response to this comment, the Rice group has
characterized their trapping potential~\cite{rice_reply}, and this
possible interpretation of their original data has now been ruled
out. Still, we want to point out (see Ref.~\cite{zwie06_comment}) that anharmonicities are of general importance for Fermi
gas experiments. These are usually carried out in traps
with trap depths comparable to the Fermi energy. Therefore,
anharmonicities have to be taken into account for any quantitative
interpretation of density profiles and accurate determination of
interaction energies from cloud sizes.

To demonstrate phase separation, i.e. the presence of a fully paired core surrounded by a shell of excess atoms, one needs to show that the
density difference of the two components abruptly changes from
zero in the superfluid core to a finite value in the normal
state. This requires stronger evidence than doubly integrated
density profiles, for example a tomographic reconstruction of the
three-dimensional density or a careful comparison with simulated
profiles.  In particular, it is necessary to distinguish between
partial and full expulsion of the unpaired majority atoms, which
was not done in~\cite{part06phase,zwie05imbalance}, but subsequently in
~\cite{shin06_prl}.  Mathematically, it is only possible to invert
one-dimensional profiles and reconstruct the 3D density, if
anharmonicities and deviations from the LDA are either absent or
fully characterized. Given the breakdown of the LDA in the data
of~\cite{part06phase}, it is mathematically impossible to infer
whether their profiles represent phase-separated clouds or only
distorted clouds. Therefore, our recent observation of the absence
of phase separation at large population
imbalances~\cite{shin06_prl} is not inconsistent with the Rice
data, but inconsistent only with their interpretation.

A breakdown of the LDA may be accentuated by phase separation, but
cannot be used as strong evidence for it.  For example, a
non-interacting Bose-Einstein condensate strongly violates the
LDA, and if it coexists with an interacting Bose-Einstein
condensate, the difference profile may show double peaks without
any interactions between the condensates.  The cross-over observed
in~\cite{part06phase} from single peak to double peak profiles may
indicate an increasing importance of LDA violation, but does not
provide evidence for a quantum phase transition.  Indeed, recent
more detailed experiments~\cite{shin06_prl} have observed shell
structures at small imbalances and seem to rule out the quantum
phase transition claimed by~\cite{part06phase}.

Distorted density profiles do not imply superfluidity. Strongly
interacting Fermi gases already show distorted profiles in the
normal phase~\cite{zwie06_nature,shin06_prl}. It is only the
observation of an abrupt change as a function of temperature that
can indicate superfluidity and allows to distinguish distortions
due to superfluidity from those due to interactions present
already in the normal phase. No abrupt changes in the density
profiles as a function of temperature were reported
in~\cite{part06phase}. It was recently found that a density change
at the phase transition occurs only as a small feature in the
clouds' center~\cite{zwie06_nature,shin06_prl}. The fact, that~\cite{part06phase} missed
the phase transition from a superfluid to a normal state at an
imbalance around 0.75, which has been seen now with three
different methods~\cite{zwie05imbalance, zwie06_nature,
shin06_prl}, implies that the observations in
Ref.~\cite{part06phase} are either not sensitive to superfluidity or that
the behavior of their sample is dominated by finite size effects.
Ref.~\cite{part06phase} emphasizes the good agreement of the cloud
size with theoretical predictions as evidence for superfluidity
(through the so-called beta factor), but the cloud size is not
expected to vary strongly as the temperature is increased above
the phase transition temperature. For an imbalanced mixture, this
was shown in~\cite{zwie06_nature}.

In conclusion, Ref.~\cite{part06phase} reports interesting results
about interaction-induced redistributions of atoms which have been linked to the breakdown of the LDA. However, the claims of the observation of pairing~\cite{part06phase}, the observation of
a superfluid core surrounded by a shell of excess spin-up
atoms~\cite{part06phase,press_release}, the observation of a
quantum phase transition~\cite{part06phase,rice_reply}, and
speculations about a new exotic form of
pairing~\cite{part06phase,press_release} are not supported by the
data.

We thank Pierbiagio Pieri, Erich Mueller and Fr\'ed\'eric Chevy for
valuable discussions.


\begin{thebibliography}{1}

\bibitem{part06phase}
G.~B. Partridge, W.~Li, R.~I. Kamar, Y.~Liao, R.~G. Hulet, {\it Science\/}
  {\bf 311}, 503 (2006). Published online 21 December 2005
  (10.1126/science.1122876).

\bibitem{desi06dens}
T.~N.~D. Silva and E.~J. Mueller, Phys. Rev. A 73, 051602 (2006).

\bibitem{haqu06dens}
M.~Haque and H.~T.~C. Stoof, Phys. Rev. A 74, 011602 (2006).

\bibitem{zwie06_comment}
M.~W. Zwierlein, W. Ketterle, Preprint cond-mat/0603489 v1

\bibitem{rice_reply}
G.~B. Partridge, W.~Li, R.~I. Kamar, Y.~Liao, R.~G. Hulet,
Preprint cond-mat/0605581.

\bibitem{zwie05imbalance}
M.~W. Zwierlein, A.~Schirotzek, C.~H. Schunck, W.~Ketterle, {\it Science\/}
  {\bf 311}, 492 (2006). Published online 22 December 2005
  (10.1126/science.1122318).

\bibitem{shin06_prl}
Y. Shin, M.~W. Zwierlein, C.~H. Schunck, A. Schirotzek, W.
Ketterle, Phys. Rev. Lett. 97, 030401 (2006).

\bibitem{zwie06_nature}
M.~W. Zwierlein, C.~H. Schunck, A.~Schirotzek, W.~Ketterle,  {\it
Nature\/} {\bf 442}, 54 (2006).

\bibitem{press_release}
Rice University, three press releases on 12/22/05, 1/12/06 and
3/14/06:
http://www.media.rice.edu/media/NewsBot.asp?MODE=VIEW\&ID=8101,
http://www.media.rice.edu/media/NewsBot.asp?MODE=VIEW\&ID=8119,
http://www.media.rice.edu/media/NewsBot.asp?MODE=VIEW\&ID=8325.



\end{thebibliography}
\end{document}